\documentclass{ws-mpla}

\def\d3k{{\displaystyle {\rm d}{\bf k} \over \displaystyle (2\pi)^3}}
\def\hmpc{h^{-1} {\rm Mpc}}

\begin{document}

\markboth{Rien van de Weijgaert}
{Cosmic Voids}

%%%%%%%%%%%%%%%%%%%%% Publisher's Area please ignore %%%%%%%%%%%%%%
\catchline{}{}{}{}{}
%%%%%%%%%%%%%%%%%%%%%%%%%%%%%%%%%%%%%%%%%%%%%%%%%%%%%%%%%%%%%%%%%%%

\title{COSMIC VOIDS:\\ STRUCTURE, DYNAMICS AND GALAXIES}

\author{\footnotesize RIEN VAN DE WEYGAERT}
\author{\footnotesize ERWIN PLATEN}

\address{Kapteyn Astronomical Institute, University of Groningen, P.O. Box 800, 
9700 AV Groningen, the Netherlands\\
weygaert@astro.rug.nl}

\maketitle

\pub{Received (28 02 2009)}{}

\begin{abstract}
In this contribution we review and discuss several aspects of Cosmic Voids. 
Voids are a major component of the large scale distribution of matter and galaxies 
in the Universe. Their instrumental importance for understanding the emergence 
of the Cosmic Web is clear. Their relatively simple shape and structure makes 
them into useful tools for extracting the value of a variety cosmic parameters, 
possibly including even that of the influence of dark energy. Perhaps 
most promising and challenging is the issue of the galaxies found within their 
realm. Not only does the pristine environment of voids provide a promising 
testing ground for assessing the role of environment on the formation and 
evolution of galaxies, the dearth of dwarf galaxies may even represent a 
serious challenge to the standard view of cosmic structure formation. 

\end{abstract}

\section{Cosmic Depressions}
\label{sec:webvoid}
\begin{figure*}
\begin{center}
\mbox{\hskip -0.18truecm\includegraphics[width=0.99\textwidth]{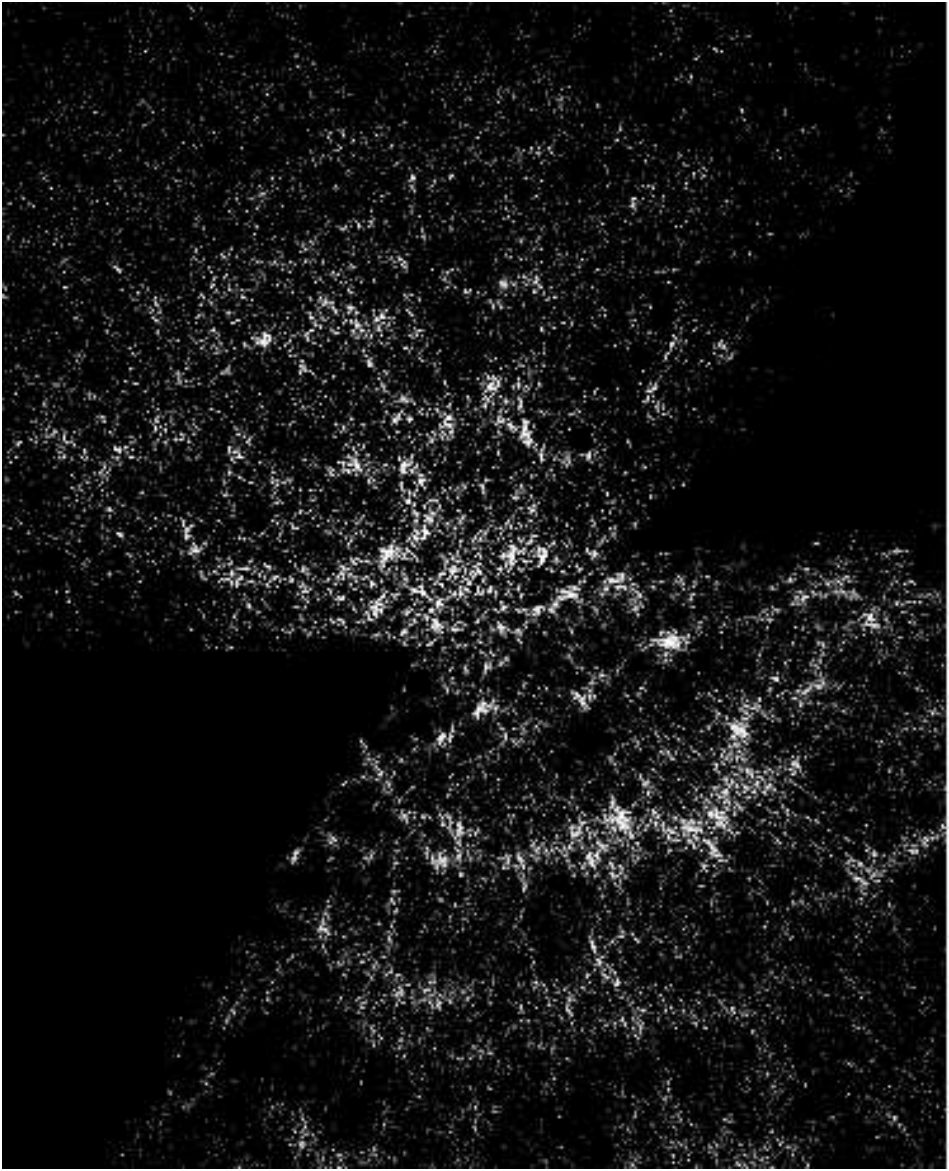}}
\end{center}
\begin{center}
\vskip 0.0truecm
\caption{SDSS is the largest and most systematic sky survey in the history of astronomy. It is a 
combination of a sky survey in 5 optical bands of 25\% of the celestial (northern) sphere. Each 
image is recorded on CCDs in these 5 bands. On the basis of the images/colours and their brightness 
a million galaxies are subsequently selected for spectroscopic follow-up. The total sky area covered 
by SDSS is 8452 square degrees. Objects will be recorded to $m_{lim}=23.1$. In total the resulting 
atlas will contain 10$^8$ stars, 10$^8$ galaxies and 10$^5$ quasars. Spectra are taken of around 10$^6$ 
galaxies, 10$^5$ quasars and 10$^5$ unusual stars (in our Galaxy). Of the 5 public data releases 4 have been 
accomplished, ie. 6670 square degrees of images is publicly available, along with 806,400 spectra. 
In total, the sky survey is now completely done (107\%), the spectroscopic survey for 68\%. This image 
is taken from a movie made by Subbarao, Surendran \& Landsberg (see website: 
http://astro.uchicago.edu/cosmus/projects/sloangalaxies/). It depicts the resulting redshift 
distribution after the 3rd public data release. It concerns 5282 square degrees and contained 
528,640 spectra, of which 374,767 galaxies.}
\end{center}
\label{fig:sdssgaldist}
\end{figure*}

Voids have been known as a feature of galaxy surveys since the first surveys were compiled \cite{chincar1975,gregthomp1978,einasto1980}.  {\it Voids} are enormous regions with sizes in the range of $20-50h^{-1}$ Mpc that 
are practically devoid of any galaxy, usually roundish in shape and occupying the major share of space in the Universe. 
Forming an essential ingredient of the {\it Cosmic Web} \cite{bondweb1996}, they are surrounded by elongated filaments, 
sheetlike walls and dense compact clusters.  Following the discovery by Ref.~\refcite{kirshner1981} and \refcite{kirshner1987} 
of the most 
dramatic specimen, the Bo\"otes void, a hint of their central position within a weblike arrangement came with the first 
CfA redshift slice \cite{lapparent1986}. This view has been dramatically endorsed and expanded by the redshift maps of 
the 2dFGRS and SDSS surveys \cite{colless2003,tegmark2004}. They have established voids as an integral component of the 
Cosmic Web. The 2dFGRS maps and SDSS maps (see e.g. fig.~\ref{fig:sdssgaldist}) are telling illustrations of the ubiquity 
and prominence of voids in the cosmic galaxy distribution.

In a void-based description of the evolution of the cosmic matter distribution, voids mark the transition scale at which 
density perturbations have decoupled from the Hubble flow and contracted into recognizable structural features. On the 
basis of theoretical models of void formation one might infer that voids may act as the key organizing element for 
arranging matter concentrations into an all-pervasive cosmic network \cite{icke1984,regoes1991,weygaertphd1991,shethwey2004}. 
As voids expand, matter is squeezed in between them, and sheets and filaments form the void boundaries. 
This view is supported by numerical studies and computer simulations of the gravitational evolution of voids in 
more complex and realistic configurations \cite{martel1990,regoes1991,dubinski1993,weykamp1993,goldvog2004,colberg2005b,padilla2005}. 
A marked example of the evolution of a typical large and deep void in a $\Lambda$CDM scenarios is given by 
the time sequence of six frames in fig.~\ref{fig:lcdmvoid}. 

Soon after their discovery, various studies pointed out their essential role in the organization of the cosmic matter 
distribution \cite{icke1984,regoes1991}. Their effective repulsive influence over their surroundings has been recognized in 
various galaxy surveys in the Local Universe. 

\subsection{Voids: cosmological importance}
There are a variety of reasons why the study of voids is interesting for our understanding of the cosmos.
\begin{enumerate}
\item[$\bullet$] They are a prominent aspect of the Megaparsec Universe, instrumental in the spatial 
organization of the Cosmic Web. 
\item[$\bullet$] Voids contain a considerable amount of information on the underlying cosmological scenario and 
on global cosmological parameters. Notable cosmological imprints are found in the outflow velocities and 
accompanying redshift distortions \cite{dekelrees1994,martel1990,rydmel1996}. Also their intrinsic structure, 
shape and mutual alignment are sensitive to the cosmology, including that of dark energy \cite{parklee2007,leepark2007,platen2008}. 
The cosmological ramifications of the reality of a supersized voids akin 
to the identified ones by ref.~\refcite{rudnick2007} and ref.~\refcite{granett2008a} would obviously be far-reaching.
\item[$\bullet$] The pristine low-density environment of voids represents an ideal and pure setting for the study 
of galaxy formation and the influence of cosmic environment on the formation of galaxies.
Voids are in particular interesing following the observation by Peebles that the dearth of 
low luminosity objects in voids is hard to understand within the $\Lambda$CDM cosmology \cite{peebles2001}. 
\end{enumerate}
\subsection{Void Sizes}
For the most systematic and complete impression of the cosmic void population the Local Universe provides the 
most accessible region. Recently, the deep view of the 2dFGRS and SDSS probes (see fig.~\ref{fig:sdssgaldist}) has been 
supplemented with high-resolution studies of voids in the nearby Universe. Based upon the 6dF survey \cite{heathjones2004}, 
Fairall (person. commun.) identified nearly all voids within the surveyed region out to $35,000$~km~s$^{-1}$. It is 
the 2MASS redshift survey \cite{huchra2005} -- the densest all-sky redshift survey available -- which has provided a 
uniquely detailed and complete census of large scale structures in our Local Universe \cite{erdogdu2006}. 

Voids in the galaxy distribution account for about 95\% of the total volume \cite{kauffair1991,eladpir1996,eladpir1997,hoyvog2002a,plionbas2002,rojas2005,platen2007}. The typical sizes of voids in the galaxy distribution depend on the galaxy population 
used to define the voids. Voids defined by galaxies brighter 
than a typical $L_*$ galaxy tend to have diameters of order $10-20h^{-1}$Mpc, but voids associated with rare luminous galaxies 
can be considerably larger; diameters in the range of $20h^{-1}-50h^{-1}$Mpc are not uncommon \cite{hoyvog2002a,plionbas2002}. 
Firm upper limits on the maximum void size have not yet been set. Recently there have been claims of the 
existence of a supersized void, in the counts of the NVVS catalogue of radio sourcs, and of its possible imprint on 
the CMB via the ISW effect in the form of a 5$^{\circ}$ CMB `Cold Spot' \cite{rudnick2007}. A systematic search for 
such supervoids, and superclusters, in the Luminous Red Galaxy (LRGs) sample from the SDSS survey has indeed provided 
one of the strongest claims for a significant (4$\sigma$) detection of the ISW effect in the WMAP maps 
\cite{granett2008a,granett2008b}. If this will be confirmed it will pose an interesting challenge 
to concordance cosmological scenarios \cite{hunt2008}. 

At the low end side of the void size distribution, a few studies claim to have found what may 
be the smallest genuine voids in existence. On the basis of the Catalog of Neighbouring Galaxies 
\cite{karachentsev2004}, ref.~\refcite{tikhklyp2007} identified a total of some 30 minivoids, each completely 
devoid of galaxies, with sizes of $0.7-3.5h^{-1}$Mpc. 

\section{Formation and Evolution of Voids}
\label{sec:voidevol}
At any cosmic epoch the voids that dominate the spatial matter distribution are a manifestation of the cosmic structure 
formation process reaching a non-linear stage of evolution.

\begin{figure*}[b]
  \center
     \vskip -7.0truecm
     \mbox{\hskip -1.1truecm\includegraphics[width=14.7cm]{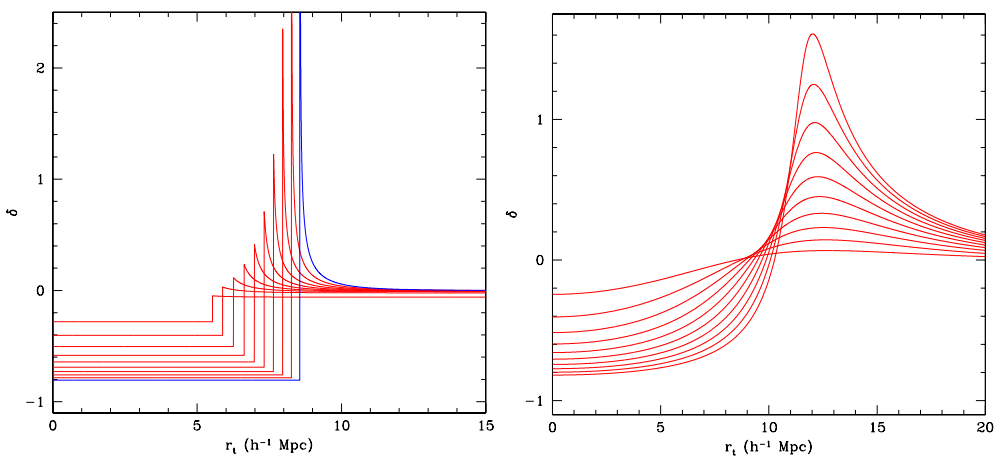}}
     \vskip -7.2truecm
\caption{Spherical model for the evolution of voids.  
 Left: a pure (uncompensated) tophat void evolving up to the epoch of 
 shell-crossing. Initial (linearly extrapolated) density deficit was 
 $\Delta_{lin,0}=-10.0$, initial (comoving) radius 
${\widetilde R}_{i,0}=5.0 h^{-1}\hbox{Mpc}$.  Right: a void with an angular averaged 
SCDM profile. Initial density deficit 
and characteristic radius are same as for  the tophat void (left). The 
tendency of this void to evolve into a tophat configuration by the time of 
shell crossing is clear. Shell-crossing, and the formation of a ridge, 
 happens only if the initial profile is sufficiently steep.}
\label{fig:sphervoid}
\end{figure*}
\begin{figure*}
\vskip -0.25truecm
  \begin{center}
     \mbox{\hskip 0.0truecm\includegraphics[height=17.5cm]{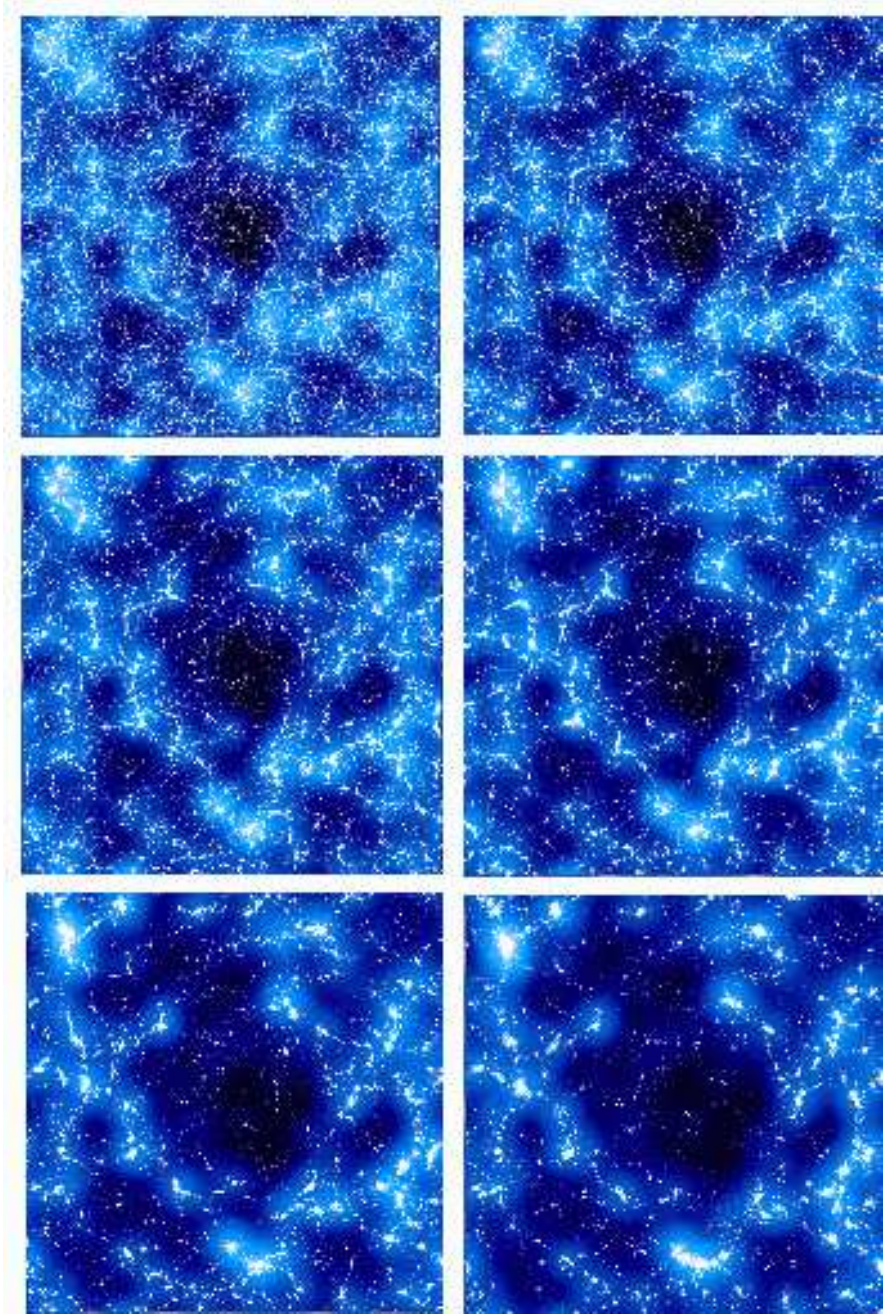}}
\caption{Simulation of evolving void (LCDM scenario). A void in a $n=0$ power-law power spectrum 
model. The slice is $50\hmpc$ wide and $10\hmpc$ thick. Shown are the partciles and smoothed 
density field (smoothed on a scale of $4\hmpc$) at six different timesteps: a=0.05, 0.15, 0.35, 
0.55, 0.75 and 1.0. Image courtesy of Erwin Platen}
\end{center}
\label{fig:lcdmvoid}
\end{figure*}

Voids emerge out of the density troughs in the primordial Gaussian field of density fluctuations. Early theoretical models 
of void formation concentrated on the evolution of isolated voids \cite{hoffshah1982,icke1984,edbert1985,blumenth1992}. 
As a result of their underdensity voids represent a region of weaker gravity, resulting in an effective repulsive peculiar 
gravitational influence. Initially underdense regions therefore expand faster than the Hubble flow, and thus expand 
with respect to the background Universe. Fig.~\ref{fig:sphervoid} illustrates the evolution of a spherical isolated 
void. As matter streams out of the void, the density within the void decreases, with isolated voids asymptotically 
evolving towards an underdensity $\delta=-1$, pure emptiness. The same expanding and evacuating behaviour of 
void regions apply in the far more complex circumstances of the real cosmic matter distribution. The illustration 
of a void in a $\Lambda$CDM Universe is illustrated in fig.~\ref{fig:lcdmvoid} by a sequence of 6 timesteps.  
Because the density within underdense regions gradually increases outward, we see a decrease of the 
corresponding peculiar (outward) gravitational acceleration: void matter in the centre moves outward faster 
than void matter towards the boundary. This leads to matter accumulating in ridges which surround the void, while 
the interior evolves into a uniform low-density 
region resembling a low-density homogeneous FRW Universe. The steepness of the resulting density profile depends on 
the protovoid depression \cite{palmvogl1983}. In nearly all conceivable situations the void therefore appears to 
assume a {\it bucket shape}, with a uniform interior density depression and a steep outer boundary 
(fig.~\ref{fig:sphervoid}, right frame). 

A characteristic evolutionary timescale for voids is that of {\it shellcrossing}. This happens when interior shells of matter 
take over initially exterior shells. Ref.~\refcite{edbert1985} demonstrated that once voids have passed the stage of shellcrossing 
they enter a phase of self-similar expansion (fig.~\ref{fig:sphervoid}). Subsequently, their expansion will slow down with 
respect to the earlier linear expansion. This impelled ref.~\refcite{blumenth1992} to identify voids in the present-day galaxy 
distribution with 
voids that have just reached the stage of shell-crossing. It happens when a primordial density depression attains a linearly 
extrapolated underdensity $\delta_v=f_v=-2.81$ (for an EdS universe, see sec~\ref{sec:voidexcur}). A perfectly 
spherical "bucket" void will have expanded by a factor of 1.72 at shellcrossing, and therefore have evolved into an underdensity 
of $\sim 20\%$ of the global cosmological density, 
ie. $\delta_{v,nl}=-0.8$. In other words, the voids that we see nowadays in the galaxy distribution do probably correspond to 
regions whose density is $\sim 20\%$ of the mean cosmic density (note that it may be different for underdensity in the 
galaxy distribution). 

Note that while by definition voids correspond to density perturbations of at most unity, $|\delta_v|\leq 1$, mature voids 
in the nonlinear matter distribution do represent {\it highly nonlinear} features. This may be best understood 
within the context of Lagrangian perturbation theory \cite{sahnshan1996}. Overdense fluctuations may be described as 
a converging series of higher order perturbations. The equivalent perturbation series is less well behaved for voids:  
successive higher order terms of both density deficit and corresponding velocity divergence alternate between negative 
and positive.

\section{Void Dynamics}
\begin{figure*}
\begin{center}
\vskip -0.0truecm
\mbox{\hskip -0.2truecm\includegraphics[width=12.1cm]{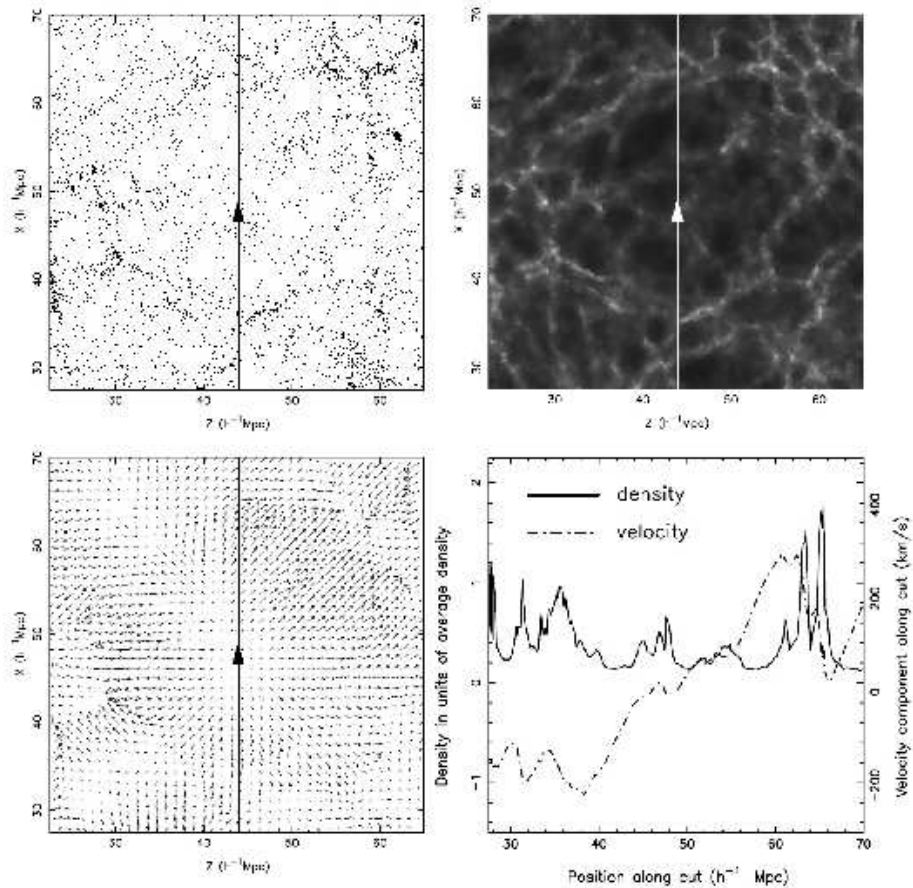}}
\vskip 0.0truecm
\caption{The density and velocity field around a void in the GIF LCDM simulation. The 
top righthand panel shows the {\rm N}-body simulation particle distribution within a slice 
through the simulation box, centered on the void. The top righthand panel shows the 
grayscale map of the DTFE density field reconstruction in and around the void, the 
corresponding velocity vector plot is shown in the bottom lefthand panel. Notice the 
detailed view of the velocity field: within the almost spherical global outflow 
of the void features can be recognized that can be identified with the diluted 
substructure within the void. Along the solid line in these panels we determined 
the linear DTFE density and velocity profile (bottom righthand frame). We can recognize 
the global ``bucket'' shaped density profile of the void, be it marked by substantial 
density enhancements. The velocity field reflects the density profile in detail, 
dominated by a global super-Hubble outflow. From Schaap 2007.}
\label{fig:gifvoid}
\end{center}
\vskip -0.5truecm
\end{figure*}

Figure~\ref{fig:gifvoid} shows a typical void-like region in a $\Lambda$CDM 
Universe. It concerns a $256^3$ particles GIF $N$-body simulation \cite{kauffmann1999}, 
encompassing a $\Lambda$CDM ($\Omega_m=0.3,\Omega_{\Lambda}=0.7,H_0=70\, {\rm km/s/Mpc}$) 
density field within a (periodic) cubic box with length $141h^{-1} 
{\rm Mpc}$ and produced by means of an adaptive ${\rm P^3M}$ $N$-body code.

The top left frame shows the particle distribution in and around the void 
within this $42.5\hmpc$ wide and $1\hmpc$ thick slice through the 
simulation box. In the same figure we include panels of the density and velocity 
field in the void, determined by means of a DTFE reconstruction 
\cite{schaapwey2000,schaapphd2007,weyschaap2009}. The void region appears as a slowly varying 
region of low density (top righthand frame). Notice the clear distinction between the 
empty(dark) interior regions of the void and its edges. In the interior of the void 
several smaller {\it subvoids} can be distinguished, with boundaries consisting of low 
density filamentary or planar structures. 

The general characteristics of the expanding void are most evident when 
following the density and velocity profile along a one-dimensional 
section through the void. The bottom-left frame of fig.~\ref{fig:gifvoid} shows 
these profiles for the linear section along the solid line indicated 
in the other three frames. The density profile does confirm to the general trend 
of low-density regions to develop a near uniform interior density surrounded by 
sharply defined boundaries. Nonetheless, we see that the void is interspersed by 
a rather pronounced density enhancement near its centre. 

The flow in and around the void is dominated by the outflow of matter from the void, 
culminating into the void's own expansion near the outer edge. The comparison 
with the top two frames demonstrates the strong relation with features in the 
particle distribution and the density field. Not only it is slightly elongated along 
the direction of the void's shape, but it is also sensitive to some prominent internal 
features of the void.  Towards the ``SE'' 
direction the flow appears to slow down near a ridge, near the centre the DTFE 
reconstruction identifies two expansion centres. 

The void velocity field profile is intimately coupled to that of its density field. 
The linear velocity increase is a manifestation of its general expansion. The near constant velocity 
divergence within the void conforms to the {\it super-Hubble flow} expected for the near 
uniform interior density distribution. Because voids are emptier than the rest of the universe 
they will expand faster than the rest of the universe with a net 
velocity divergence equal to
\begin{eqnarray}
   \theta&\,=\,&{\displaystyle \nabla\cdot{\bf v} \over \displaystyle H}\,=\,3 (\alpha-1)\,,\qquad\alpha=H_{\rm void}/H\,,\\
\end{eqnarray}
\noindent where $\alpha$ is defined to be the ratio of the super-Hubble expansion rate of the 
void and the Hubble expansion of the universe.

\subsection{Dynamical Influence}
Various studies have found strong indications for the imprint of voids in the peculiar velocity flows of 
galaxies in the Local Universe. Ref.~\refcite{bothun1992} made the first claim of seeing pushing influence of voids 
when assessing the stronger velocity flows of galaxies along a filament in the first CfA slice. Stronger 
evidence came from the extensive and systematic POTENT analysis of Mark III peculiar galaxy velocities 
\cite{willick1997} in the Local Universe \cite{dekel1990,edbert1990}. POTENT found that for a fully selfconsistent 
reconstruction of the dynamics in the Local Universe, it was inescapable to include the dynamical influence of voids 
\cite{dekel1994}. 

\begin{figure*}[t]
  \vskip -0.5truecm
  \centering
    \mbox{\hskip -0.5truecm\includegraphics[width=9.5cm,angle=270.0]{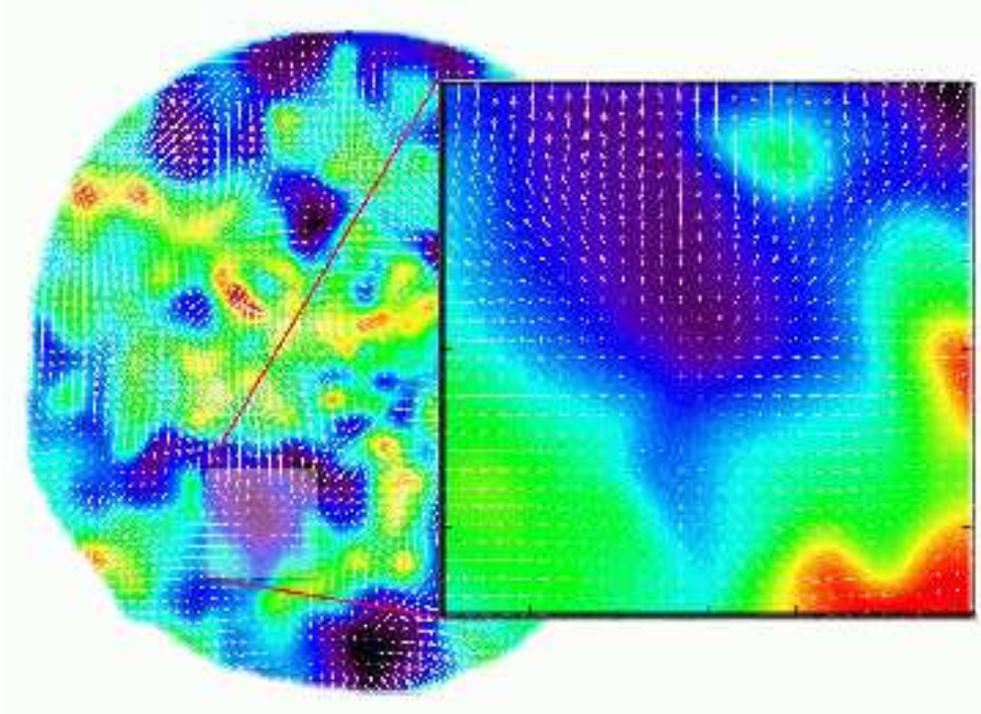}}
    \vskip -0.0truecm
%\end{figure*}
%\begin{figure*}[t]
    \caption{Gravitational impact of the Sculptor Void. The righthand frame shows the inferred velocity field 
    in and around the Sculptor void near the Local Supercluster. The colour map represents the density values, 
    with dark blue at $\delta \sim -0.75$ and cyan near $\delta \sim 0.0$. The vectors show the 
    implied velocity flow around the void, with a distinct nearly spherically symmetric outflow. 
    It is a zoom-in onto the indicated region in the density and velocity map in the Local Universe (lefthand) 
    determined on the basis of the PSCz galaxy redshift survey. From: Romano-D\'{\i}az \& van de Weygaert  2007.
} 
\label{fig:psczsculptor} 
\end{figure*} 
With the arrival of new and considerably improved data samples the dynamical influence of voids in the 
Local Universe has been investigated and understood in greater detail. The reconstruction of the density 
and velocity field in our local cosmos on the basis of the 2MASS redshift survey has 
indeed resulted in a very interesting and complete view of the dynamics on Megaparsec scales.  
This conclusion agree with that reached on the basis of an analysis of the 
peculiar velocity of the Local Group by ref.~\refcite{tully2008}. Their claim is that the Local Void is 
responsible for a considerable repulsive influence, accounting for $\sim 259$~km~s$^{-1}$ of the 
$\sim 631$~km~s$^{-1}$ Local Group motion with respect to the CMB. 

\begin{figure*}
\begin{center}
\vskip -0.0truecm
\mbox{\hskip -0.0truecm\includegraphics[width=12.1cm]{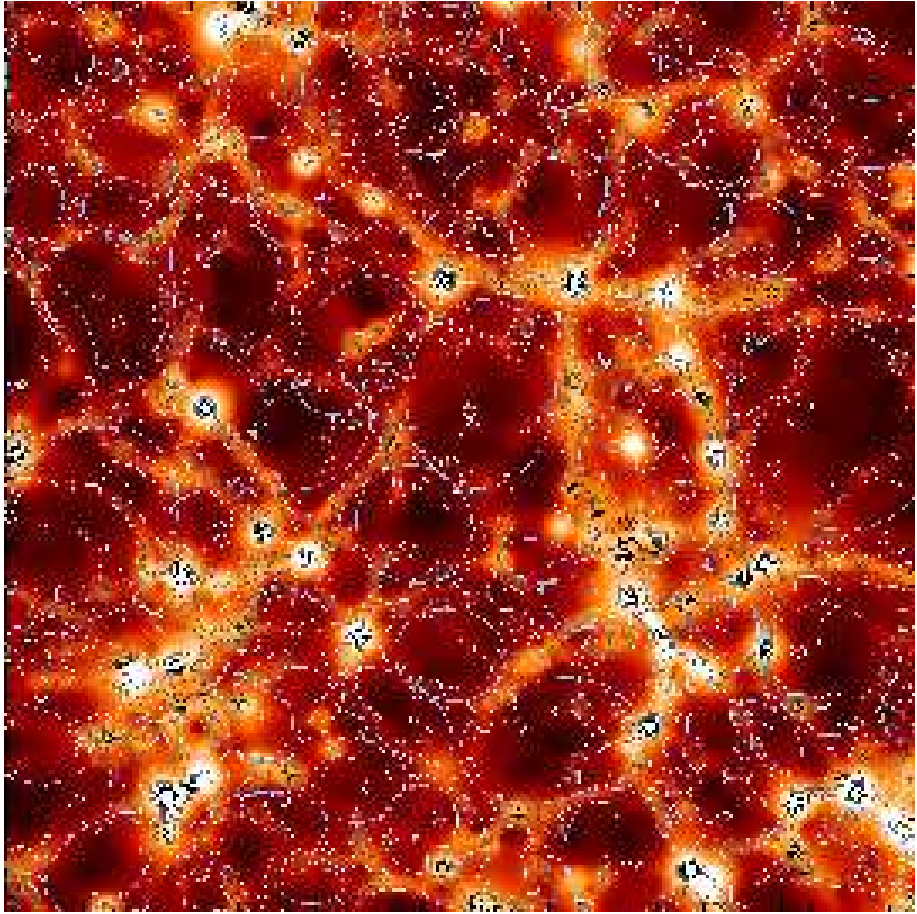}}
\vskip 0.0truecm
\caption{Voids and Void Dynamics in a $\Lambda$CDM cosmology. In a 100$\hmpc$ box, 
the $1 \hmpc$ Gaussian smoothed density field contour levels are superimposed on the 
$1 \hmpc$ smoothed map of the tidal force field strength (note the outstanding cluster 
locations), along with ellipses identifying the voids by size, shape and orientation 
(on a scale of $6 \hmpc$).}
\label{fig:voidtide}
\end{center}
\vskip -0.5truecm
\end{figure*}

\subsection{Void Shapes and Tides}
Ref.~\refcite{icke1984} pointed out that any (isolated) aspherical underdensity will become more spherical as it expands. 
The effective gravitational acceleration is stronger along the short axis than along the longer axes. For 
overdensities this results in a stronger inward acceleration and infall, producing increasingly flattened 
and elongated features. By contrast, for voids this translates into a larger {\it outward} acceleration 
along the shortest axis so that asphericities will tend to diminish. For the interior of voids this 
tendency has been confirmed by {\rm N}-body simulations \cite{weykamp1993}. 

In reality, voids will never reach sphericity. Even though voids tend to be less flattened or elongated than 
the halos in the dark matter distribution, they are nevertheless quite nonspherical: ref.~\refcite{platen2008} find 
that they are slightly prolate with axis ratios of the order of $c:b:a\approx 0.5:0.7:1$. This agrees 
with the statement by ref.~\refcite{shandarin2006} and ref.~\refcite{parklee2007} that in realistic cosmological circumstances 
voids will be nonspherical. This is also quite apparent in images of, for example, the Millennium simulation 
\cite{springmillen2005}. 

The flattening is a result of large scale dynamical and environmental factors, amongst which we can identify at 
least two important factors \cite{platen2008}. Even while their internal dynamics pushes them to a more 
spherical shape they will never be able to reach perfect sphericity before encountering surrounding structures such as 
overdense filaments or planar walls. Even more important may be the fact that, for voids, external tidal
influences remain important: voids will always be rather moderate densities perturbations since they can 
never be more underdense than $\delta=-1$. External tidal forces are responsible for a significant 
anisotropic effect in the development of the voids. In extreme cases they may even cause complete collapse 
of the void.   

\begin{figure*}[b]
\begin{center}
\vskip -3.0truecm
\mbox{\hskip -0.5truecm\includegraphics[width=13.4cm]{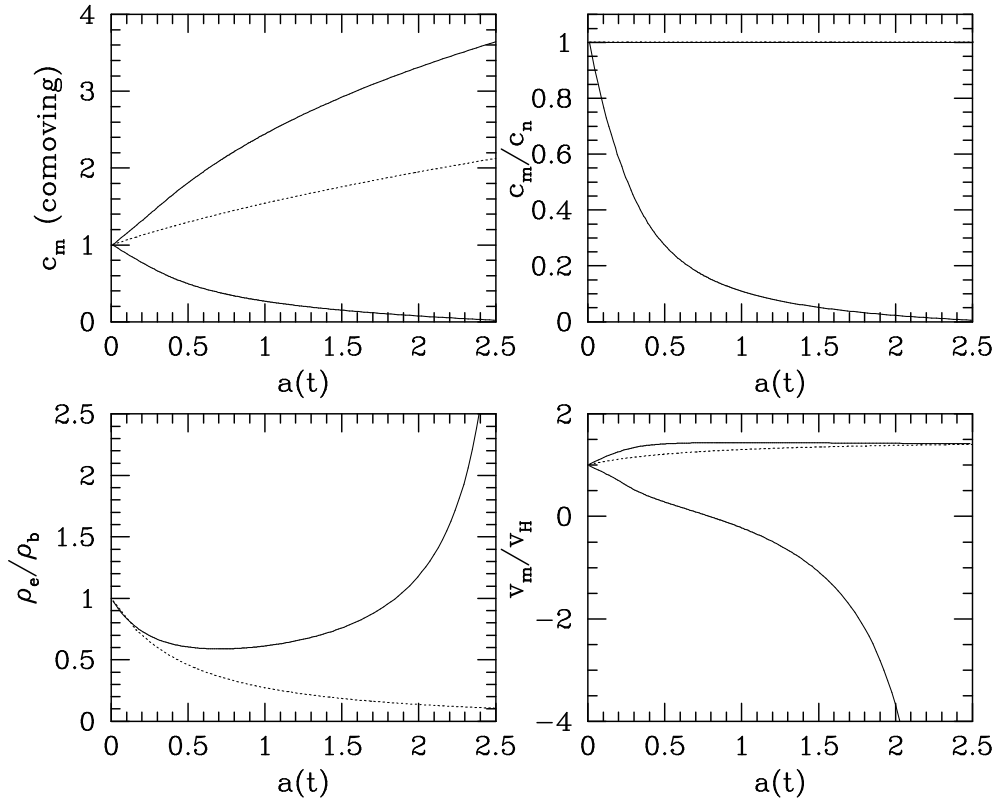}}
\vskip -0.25truecm
\caption{Anisotropic Void Expansion and Collapse. Evolution of the 
same initially spherical void. Dashed line: without external influences  
Solid line: under the influence of an axisymmetric and linearly evolving 
external tidal force field ($E_{mm,0}=(-E,-E,2E)$).
{\it Topleft:} (comoving) void axis, $c_m'=c_m/a(t)$. 
{\it Topright:} axis ratios $c_2/c_1$ and $c_3/c_1$. 
{\it Bottom Left:} Internal density $\rho_e$ of the void, in 
units of the cosmic density $\rho_u(t)$. {\it Bottom Right:} 
The velocity $v_m$ along the axes of the voids, in units of the 
Hubble velocity $v_H$. Note the collapse along axis 1 and 2.}
\label{fig:voidcoll}
\end{center}
\end{figure*}

\subsubsection{Ellipsoidal Voids}
The major influence of external tidal forces on the evolution of voids can be 
understood by assessing the evolution of homogeneous underdense ellipsoids \cite{icke1984}.
To some extent, the {\it Homogeneous Ellipsoidal Model} \cite{icke1973,whitesilk1979,eisenstloeb1995,bondmyers1996,desjacques2007} 
represents a better description 
for the evolution of voids than that of collapsing overdense peaks. Overdense regions contract into more 
compact and hence steeper density peaks, so that the area in which the ellipsoidal model represents 
a reasonable approximation will continuously shrink. On the other hand, we have seen the natural 
tendency of voids is to develop an increasingly uniform interior density field (see 
fig.~\ref{fig:sphervoid}): while they expand their interior gets drained of matter and 
develops a flat ``bucket-shaped'' density profile.

In many respects the homogeneous model is a better approximation for underdense regions than it is for 
overdense ones. Evidently, the approximation is restricted to the interior and fails 
at the void's outer fringes because of its neglect of the domineering role of surrounding material, 
such as the sweeping up of matter and the encounter with neighbouring features. 

The homogeneous ellipsoidal model assumes an object to be a region with a triaxially symmetric 
ellipsoidal geometry and a homogeneous interior density, embedded within a uniform background 
density $\rho_{\rm u}$. Consider the simple situation of the external tidal shear 
directed along the principal axes of the ellipsoid. The gravitational acceleration along 
the principal axes of an ellipsoid with over/underdensity $\delta$ can be evaluated from the 
expression for the corresponding scale factors ${\cal R}_{\rm i}$,
\begin{eqnarray}
{\displaystyle d^2 {\cal R}_{\rm m} \over \displaystyle d t^2}&\,=\,&- 4 \pi G \rho_{\rm u}(t)\, 
\left[{1+\delta\over 3}\,+\,{1\over 2}\,(\alpha_m-{2 \over 3})\,\delta\right]\,{\cal R}_{\rm m}\,-\,\tau_{\rm m}\,
{\cal R}_{\rm m}\,+\,\Lambda R_m\,,
\label{eq:ellipseqnmot}
\end{eqnarray}
where we have also taken into account the influence of the cosmological constant $\Lambda$. The factors 
$\alpha_{\rm m}(t)$ are the ellipsoidal coefficients specified by the integral equation,  
\begin{eqnarray}
\alpha_{\rm m}(t)\,=\,{\cal R}_1(t) {\cal R}_2(t) {\cal R}_3(t)\,{\int}^\infty_0\,
{\displaystyle {\rm d}\lambda\ \over \displaystyle ({\cal R}_{\rm m}^2(t)+\lambda )\,
\prod\nolimits_{n=1}^3 \left({\cal R}_{\rm n}^2(t)+\lambda\right)^{1/2}}\,.
\label{eq:ellalph}
\end{eqnarray}

\bigskip
\noindent The influence of the external (large-scale) tidal shear tensor $T^{(ext)}_{\rm mn}$ enters via the 
eigenvalue $\tau_m$. From eqn.~\ref{eq:ellipseqnmot}, it is straightforward to appreciate that 
as $\delta$ grows strongly nonlinear, the relative influence of the large-scale (near-)linear 
tidal field will decline. However, the density deficit $\delta$ of voids will never exceed 
unity, $|\delta|<1$, so that the importance of the factor $\tau_m$ remains relatively large. 
Its impact can be so strong that it not only effects an anisotropic expansion of the 
void, but even may manage to make an initially spherically expanding void to collapse 
(see fig.~\ref{fig:voidcoll}).

\subsubsection{Void Alignments}
Large scale tidal influences not only manifest themselves in the shaping of individual voids. They are 
also responsible for a distinct alignment of substructures along a preferred direction, while they 
are also instrumental in their mutual arrangement and organization. Locally, the orientation of a void 
turns out to be strongly aligned with the tidal force field generated by structures on scales up to at 
least $20-30h^{-1}\hbox{\rm Mpc}$. This goes along with a similar mutual alignment amongst voids themselves. 
They have strongly correlated orientations over distances $>30h^{-1}\hbox{Mpc}$ \cite{platen2008}, a scale 
considerably exceeding the typical void size. It forms a strong confirmation of the large scale 
tidal force field as the dominant agent for the evolution and spatial organization of the 
Megaparsec Universe, as emphasized by the Cosmic Web theory \cite{bondweb1996}. 

\section{Void Sociology}
\label{sec:voidsocio}
Computer simulations of the gravitational evolution of voids in realistic cosmological environments do 
show a considerably more complex situation than that described by idealized spherical or ellipsoidal models 
\cite{martel1990,regoes1991,dubinski1993,weykamp1993,goldvog2004,colberg2005b,padilla2005}. In recent years 
the huge increase in computational resources has enabled {\rm N}-body simulations to resolve in detail the intricate 
substructure of voids within the context of hierarchical cosmological structure formation scenarios 
\cite{mathis2002,gottloeb2003,hoeft2006,arbabmuell2002,goldvog2004,colberg2005b,padilla2005}. They confirm the 
theoretical expectation of voids having a rich substructure as a result of their hierarchical buildup 
(see e.g. fig.~\ref{fig:lcdmvoid}). 

Ref.~\refcite{shethwey2004} treated the emergence and evolution of voids within the context of {\it hierarchical} 
gravitational scenarios. It leads to a considerably modified view of the evolution of voids. The role of 
substructure within their interior and the interaction with their surroundings turn out to be essential aspects of 
the {\it hierarchical} evolution of the void population in the Universe. An important guideline are the
heuristic void model simulations by ref.~\refcite{dubinski1993}, and the theoretical void study by 
ref.~\refcite{sahni1994} within the context of a Lagrangian adhesion model approach by ref.~\refcite{sahni1994}. 
As argued by ref.~\refcite{shethwey2004}, the hierarchical development  of voids, akin to the evolution of overdense halos, 
may be described by an {\it excursion set} formulation \cite{pressschecht1974,bond1991,sheth1998}. 

In some sense voids have a considerably more complex evolutionary path than overdense halos. Two 
processes dictate the evolution of voids: their {\it merging} into ever larger voids as well as the {\it collapse} 
and disappearance of small ones embedded in overdense regions (see fig.~\ref{fig:voidhier}).

\subsubsection{Void Merging}
First, consider a small region which was less dense than the critical void density value.  It may be that 
this region is embedded in a significantly larger underdense region which is also less dense than the critical density. 
Many small primordial density troughs may exist within the larger void region. Once small density depressions located 
within a larger embedding underdensity have emerged as true voids at some earlier epoch, their expansion tends 
to slow down. 

\begin{figure*}[t]
\begin{center}
\vskip -0.0truecm
\mbox{\hskip -0.5truecm\includegraphics[width=13.4cm]{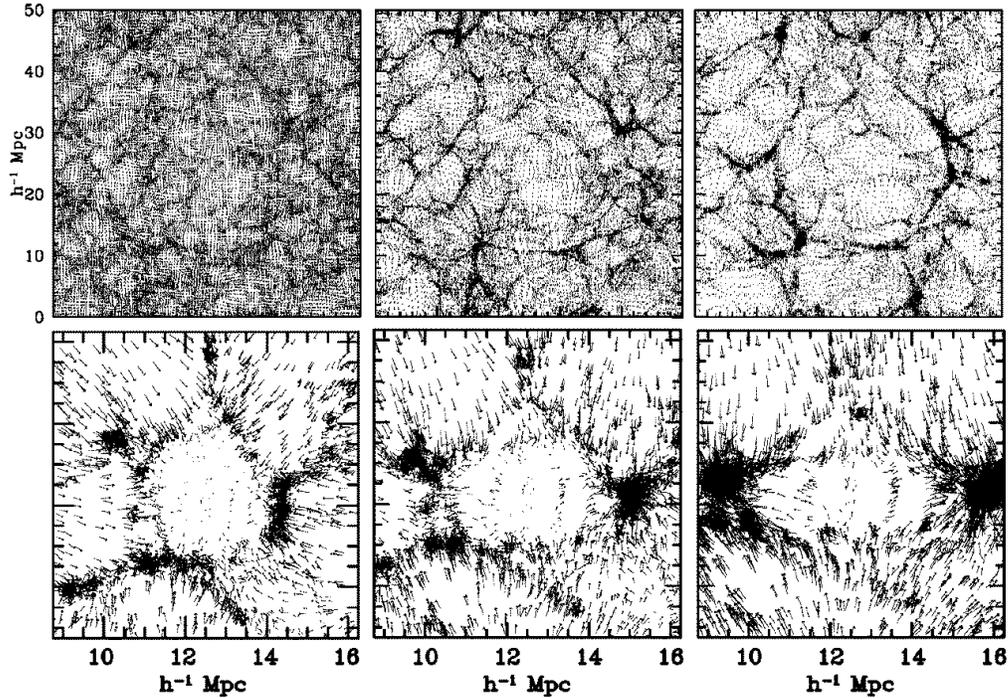}}
\vskip 0.0truecm
\caption{The two modes of void evolution: void merging (top row) and void collapse (bottom row). 
Top: three timesteps of evolving void structure in a 128$^3$ particle {\rm N}-body simulation of structure 
formation in an SCDM model ($a_{\rm exp}=0.1, 0.3,0.5$). The sequence shows the gradual development 
of a large void of diameter $\approx 25h^{-1}\hbox{Mpc}$ as the complex pattern of smaller voids and 
structures which had emerged within it at an earlier time, merge with one another. It illustrates 
the {\it void-in-void} process of the evolving void hierarchy. Bottom: a choice of three collapsing 
voids in a constrained {\rm N}-body simulation, each embedded within an environment of different tidal 
shear strength. The arrows indicate the velocity vectors, showing the infall of outer regions onto 
the void region. As a result the voids will be crushed as the surrounding matter rains down on them.}
\label{fig:voidhier}
\end{center}
\vskip -0.5truecm
\end{figure*}

When the adjacent subvoids meet up, the matter in between is squeezed in thin walls and filaments. The peculiar 
velocities perpendicular to the void walls are mostly suppressed so that the flow of matter is mostly 
confined to tangential motions. The subsequent merging of the subvoids is marked by the gradual fading of 
these structures while matter evacuates along the walls and filaments towards the enclosing boundary of 
the emerging void \cite{dubinski1993} (also see top row fig.~\ref{fig:voidhier}). The timescale on which 
the internal substructure of the encompassing void is erased is approximately the same as that on which it 
reaches maturity.

The final result is the merging and absorption of the subvoids in the larger void emerging from the embedding 
underdensity. Hence, as far as the void population is concerned only the large void counts, while the 
smaller subvoids should be discarded as such. Only a faint and gradually fading imprint of their original outline 
remains as a reminder of the initial internal substructure. 

\subsubsection{Void Collapse} 
A {\it second} void process is responsible for the radical dissimilarity between void and halo populations. If a small 
scale minimum is embedded in a sufficiently high large scale density maximum, then the collapse of the larger surrounding 
region will eventually squeeze the underdense region it surrounds: the small-scale void will vanish when the region around 
it has collapsed completely. Alternatively, though usually coupled, they may collapse as a result of the 
tidal force field in which they find themselves. If the void within the contracting overdensity has been 
squeezed to vanishingly small size it should no longer be counted as a void (see fig.~\ref{fig:voidhier}, bottom row). 

The collapse of small voids is an important aspect of the symmetry breaking between underdensities and 
overdensities. In the primorial Universe, Gaussian primordial conditions involve a perfect symmetry between under- and 
overdense. Any inspection of a galaxy redshift map or an {\rm N}-body simulation shows that there is a marked difference 
between matter clumps and voids. While the number density of halos is dominated by small objects, void collapse 
is responsible for the lack of small voids.

\subsection{Void Excursions}
\label{sec:voidexcur}
Marked by the two processes of merging and collapse, the complex hierarchical buildup of the 
void population may be modelled by a two-barrier excursion set formalism \cite{shethwey2004}. 
The barriers refer to the critical (linear) density thresholds involved with the {\it merging} 
and {\it collapse} of voids. Whenever the linearly extrapolated (primordial) $\delta_L({\vec r},t|R)$ 
on a scale $R$, 
\begin{equation}
\delta_L(r,t|R)\,=\,{\displaystyle D(t) \over \displaystyle D(t_i)}\,\delta_L(r,t_i|R)\,,
\end{equation}
exceeds a density threshold $f_c$ it will collapse. For an Einstein-de Sitter $\Omega_m=1$ Universe the 
critical value has the well-known value $f_c\,\simeq\,1.686$ \cite{gunngott1972}. A void will form 
when an underdensity reaches the critical density threshold of {\it shell-crossing}, corresponding to a 
value of $f_v\,\simeq\,-2.81$ for spherical voids in an Einstein-de Sitter Universe. The linear density growth factor 
$D(t)$, normalized to unity at the present epoch, follows from the integral \cite{heath1977,peebles1980,hamilton2001,lahavsuto2004},
\begin{eqnarray}
D(t)\,=\,D(t,\Omega_{m,0},\Omega_{\Lambda,0})\,=\,{\displaystyle 5\,\Omega_{m,0} H_0^2 \over \displaystyle 2}\,H(a)\,\int_0^a\,{\displaystyle da' \over \displaystyle a'^3 H^3(a')}\,.
\label{eq:lingrowthlambd}
\end{eqnarray}

\noindent Emerging from a primordial Gaussian random field, many small voids may coexist within one larger void. 
Small voids from an early epoch will merge with one another to form a larger void at a later 
epoch. The excursion set formalism takes account of this {\em void-in-void} configuration by discarding 
these small voids from the void count. To account for the impact of voids disappearing when embedded in 
collapsing regions, the two-barrier formalism also deals with the {\it void-in-cloud} problem. 

By contrast, the evolution of overdensities is governed only by the {\it cloud-in-cloud} process; 
the {\it cloud-in-void} process is much less important, because clouds which condense in a large scale void are not torn 
apart as their parent void expands around them.  This asymmetry between how the surrounding environment affects halo and 
void formation is incorporated into the {\it excursion set approach} by using one barrier to model halo formation and 
a second barrier to model void formation. Only the first barrier matters for halo formation, 
but both barriers play a role in determining the expected abundance of voids. 

This image of void formation in the dark matter distribution has been elaborated by 
ref.~\refcite{furlpir2006} to describe the implications for voids in the galaxy distribution.

\subsection{Voidpatch Theory}
The extended Press-Schechter formulation of hierarchical void evolution is a local description 
and fails to take into account the proper spatial and geometric setting and configuration 
underlying the process. 

The {\it peakpatch} formalism introduced by ref.~\refcite{bondmyers1996} represents a more proper 
nonlocal description of the hierarchical buildup of structures. Framed in an elaborate 
analytical framework it involves a natural definition of a peaks physical extension, 
and takes into the account of the anisotropic nature of gravitational collapse and of 
the relative location and size of peaks in and around evolving density peaks. 

Along similar lines ref.~\refcite{platen2009} have phrased a {\it voidpatch} formalism to follow the 
hierarchical buildup of voids. It will allow a much better understanding of the configuration, shape 
and substructure of individual voids emerging in a cosmic density field. It will also help 
towards the identification of collapsing subvoids and help towards evaluating the impact 
of the various dynamical influences.

\begin{figure*}
  \vskip -2.0truecm
  \begin{center}
     \mbox{\hskip -0.25truecm\includegraphics[width=12.5cm]{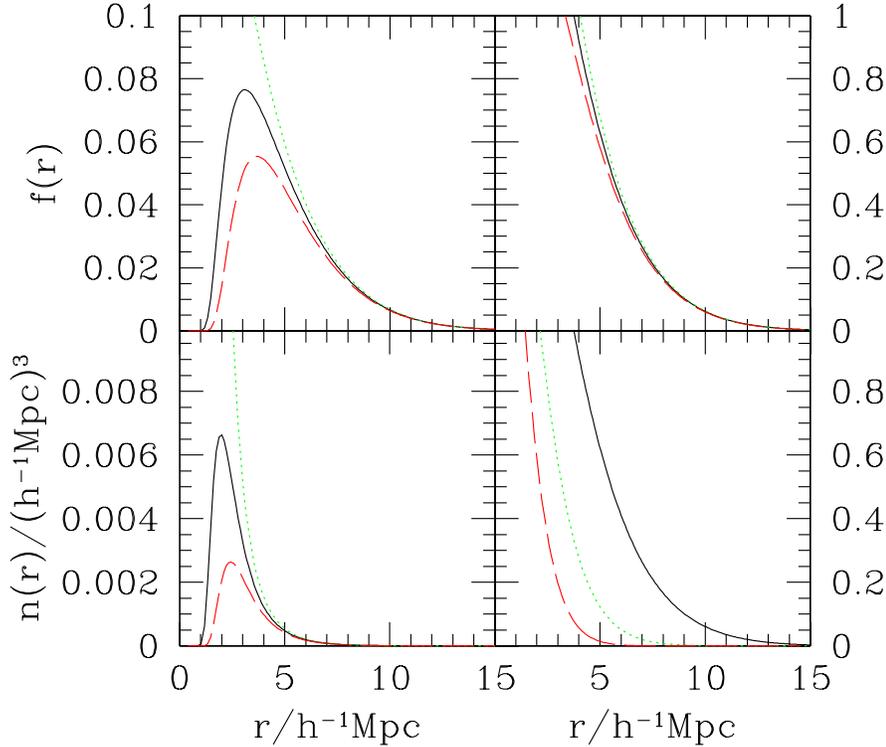}}
  \end{center}
  \vskip -1.0truecm
 \caption{Distribution of void radii predicted by the two-barrier 
  extended PS formalism, in an Einstein de-Sitter model with 
 $P(k)\propto k^{-1.5}$, normalized to $\sigma_8=0.9$ at $z=0$.  
 Top left panel shows the mass fraction in voids of radius $r$.  
 Bottom left panel shows the number density of voids of radius $r$.  
 Note that the void-size distribution is well peaked about a 
 characteristic size provided one accounts for the void-in-cloud 
 process.  Top right panel shows the cumulative distribution of the 
 void volume fraction. 
 Dashed and solid curves in the top panels and bottom left panel 
 show the two natural choices for the importance of the void-in-cloud 
 process discussed in the text:  $\delta_{\rm c}=1.06$ and 1.686, 
 with $\delta_{\rm v}=-2.81$.  Dotted curve shows the result of 
 ignoring the {\em void-in-cloud} process entirely.
 Clearly, the number of small voids decreases as the ratio of 
 $\delta_{\rm c}/|\delta_{\rm v}|$ decreases. 
 Bottom right panel shows the evolution of the cumulative void volume 
 fraction distribution. The three curves in this panel are for 
 $\delta_{\rm c}=1.686(1+z)$, where $z=0$ (solid), 0.5 (dotted) 
 and~1 (dashed). From Sheth \& van de Weygaert 2004.}
\label{fig:voidhierdist}
\end{figure*}

\subsection{Void Population Statistics.}
The analytical evaluation of the two-barrier random walk problem in the extended Press-Schechter approach leads 
directly to a prediction of the distribution function $n_v(M)$ for voids on a mass scale $M$ (or corresponding void 
size $R$\footnote{The conversion of the void mass scale to equivalent void radius $R$ is done by assuming the simplest 
approximation, that of the spherical tophat model. According to this model a void has expanded by a factor of 1.7 by 
the time it has mature, so that $V_v=(M/\rho_u)*1.7^3$.}). The resulting void spectrum\footnote{Note that for 
near-empty voids the mass scale $M$ is nearly equal to the corresponding void mass deficit.}. is peaked, with a sharp 
cutoff at both small and large values of the peak mass $M_{v,\ast}$ (fig.~\ref{fig:voidhierdist}, bottom lefthand frame),
\begin{eqnarray}
n_v(M)\,{\rm d}M&&\,\approx\nonumber\\
\ \\
&&\sqrt{\frac{2}{\pi}}\,\frac{\rho_u}{M^2}\,\nu_v(M)\,\exp\left(-\frac{\nu_v(M)^2}{2}\right)\,
\left|\frac{{\rm d} \ln \sigma(M)}{{\rm d} \ln M}\right|\,\exp\left\{-{|f_{\rm v}|\over f_{\rm c}}\,
 {{\cal D}^2\over 4\nu_v^2}-2{{\cal D}^4\over\nu_v^4} \right\}\,,\nonumber
\end{eqnarray}
where $\sqrt{\nu_v(M)}$ is the fractional relative underdensity, 
\begin{equation}
\nu_v(M)\,\equiv\,\frac{{|f_{\rm v}|}}{\sigma(M)}\,,
\label{eq:numvoid}
\end{equation}
\noindent in which with the dependence on the mass scale $M$ entering via the rms density fluctuation on that scale, 
$\sigma(M)$. The quantity ${\cal D}$ is the ``{\it void-and-cloud parameter}'', ${\cal D} \equiv |f_{\rm v}|/(f_{\rm c}+|f_{\rm v}|)$.  
It parameterizes the impact of halo evolution on the evolving population of voids: the likelihood of smaller 
voids being crushed through the {\it void-in-cloud} process decreases as the relative value of the collapse barrier 
$f_{\rm c}$ with respect to the void barrier $f_{\rm v}$ becomes larger. 

From fig.~\ref{fig:voidhierdist} we also see that the population of large voids is insensitive to the {\it void-in-cloud} 
process. The large mass cutoff of the void spectrum is similar 
to the ones for clusters and reflects the Gaussian nature of the fluctuation field from which the 
objects have condensed. The {\em characteristic void size} increases with time: the gradual merging of voids 
into ever larger ones is embodied in a self-similar shift of the peak of the void spectrum.

When evaluating the corresponding fraction of contained in voids on mass scale M, $f(M)\,=\,{M\,n_v(M)/\rho_u}\,$ 
(top lefthand frame, fig.~\ref{fig:voidhierdist}), we see that this also peaks at the characteristic 
void scale. It implies a mass fraction in voids of approximately thirty percent of the mass in the 
Universe, with most of the void mass to be found in voids of this characteristic mass. 

While the two-barrier excursion set formalism offers an attractive theoretical explanation for the 
distinct asymmetry between clumps and voids and for the peaked void size distribution, realistic 
cosmological simulations are needed to identify where the disappearing small-scale voids are to be found 
in a genuine evolving cosmic matter distribution. Using the GIF {\rm N}-body simulations of various $CDM$ scenarios, 
ref.~\refcite{platen2005} has managed to trace various specimen of this unfortunate void population. Subvoids in 
the interior of a larger void tend to merge with surrounding peers while the ones near the 
boundary get squeezed out of existence. These do not collapse isotropically, but tend to get 
sheared by their surroundings. 

An important aspect of the implied void population is that it is approximately {\it space-filling}. It underlines 
the adagio that the large scale distribution of matter may be compared to a {\it soapsud of expanding bubbles}. 
This follows from evaluation of the cumulative integral 
\begin{equation}
{\cal F}_V(M)\,\equiv\,\int_M^{\infty}\,(1.7)^3\,{\frac{M'\,n_v(M')}{\rho_u}}\,dM'\,\,.   
\end{equation}
where the factor 1.7 is an estimate of the excess expansion of the void based upon the spherical model for void 
evolution (see footnote). The top righthand panel of fig.~\ref{fig:voidhierdist} shows the resulting (current) cumulative 
void volume distribution: for a finite value of void radius $R$ the whole of space indeed appears to be occupied by voids, 
while the bottom righthand frame shows the gradual shift of the cumulative volume distribution towards larger voids. 
In other words, the correct image appears to be that of a gradually unfolding bubbly universe in which the 
average size of the voids grows as small voids merge into ever larger ones. 

\subsection{Void Substructure}
An important issue within the hierarchically proceeding evolution of voids and the Cosmic Web 
is the fate of its substructure. In voids the diluted and diminished infrastructure remains 
visible, at ever decreasing density contrast, as cinders of the earlier phases of the {\it void 
hierarchy} in which the substructure stood out more prominent

{\rm N}-body simulations show that voids do retain a rich infrastructure. Examples such as the 
top row of fig.~\ref{fig:voidhier}, and images of the Millennium simulation \cite{springmillen2005}  
show that while void substructure does fade, it does not disappear. We may find structures ranging from 
filamentary and sheetlike structures to a population of low mass dark matter halos and galaxies 
(see e.g. ref.~\refcite{weykamp1993} \& \refcite{gottloeb2003}). Although challenging, these may also be 
seen in the observational reality. The galaxies that populate the voids do currently attract quite some 
attention (see next section). Also, the SDSS galaxy survey has uncovered a substantial level of 
substructure within the Bo\"otes void, confirming tantative indications for a filamentary 
feature by ref.~\refcite{szomoru1996}.

The gradual fading of the initially rich substructure as a void expands, and its matter 
content evacuates its interior, is clearly visible in the development of the void region in 
fig.~\ref{fig:lcdmvoid}. This may be seen as the slowing of structure growth in the 
low-density environment of voids. In their analytical treatment, ref.~\refcite{goldvog2004} described 
how the evolving substructure in a void could be translated into structure evolving in a lower 
$\Omega$ Universe, including the implicit change in spectral characteristics of the density 
field (see ref.~\refcite{martino2009} for a reappraisal and criticism of some of the technical issues). 

The resulting observation is one in which the density distribution in a void region 
undergoes a manifest and continuous transformation towards an ever larger dominating scale, 
while it mostly retains the topological characteristics of the initial density field. 
In a sense, it reflects of the buildup of the Cosmic Web itself, its basic pattern 
already recognizable in the primordial matter distribution \cite{bondweb1996}.

\section{Void Galaxies}
\label{sec:voidgal}
A major point of interest is that of the galaxies populating the voids, the {\it void galaxies}. The pristine environment of 
voids represents an ideal and pure setting for the study of galaxy formation. Largely unaffected by the complexities 
and processes modifying galaxies in high-density environments, the isolated void regions must hold important clues to 
the formation and evolution of galaxies. This has made the study of the relation between void galaxies and their 
surroundings an important aspect of the recent interest in environmental influences on galaxy formation  
\cite{szomoru1996,kuhn1997,popescu1997,karachentseva1999,groggell1999,groggell2000,hoyvog2002a,hoyvog2002b,rojas2004,rojas2005,tikhonov2006,patiri2006a,patiri2006b,ceccar2006,wegner2008,stanonik2009a}. 

Amongst the issues relevant for our understanding of galaxy and structure formation, void galaxies have posed several 
interesting riddles and questions. Arguably the most prominent issue is that of the near absence of low-luminosity 
galaxies in voids, while $\Lambda$CDM cosmology expects voids to be teeming with dwarfs and low surface brightness 
galaxies. A manifest environmental influence which still needs to be understood is the finding that void galaxies appear 
bluer, a trend which continues down into the most rarefied void regions. Another issue of interest is whether we can 
observe the intricate filigree of substructure in voids, expected as the remaining debris of the merging voids 
and filaments in the hierarchical formation process. 

\begin{figure}[t]
  \centering
  \includegraphics[width=1.0\linewidth]{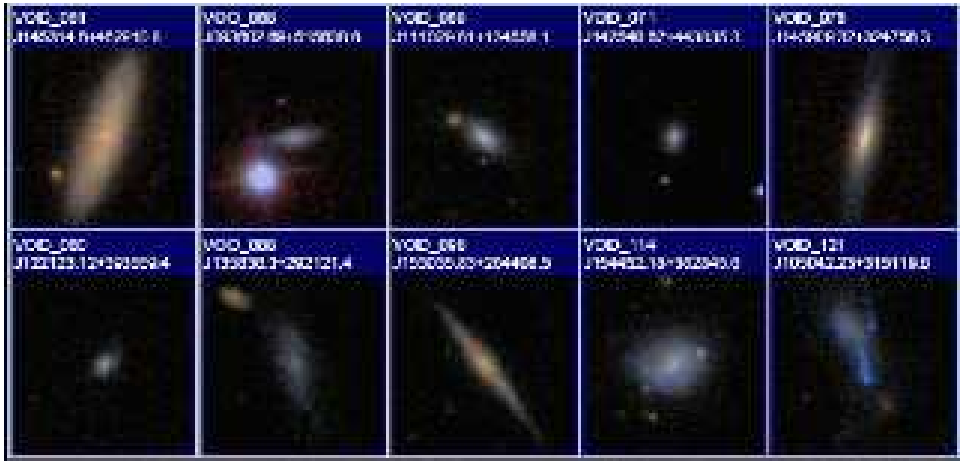}
  \caption{\small A selection of 10 void galaxies from the SDSS DR7 galaxy redshift survey. These 
galaxies are part of the WSRT HI void galaxy survey. Courtesy: Arag\'on-Calvo, Platen, van de Weygaert.}
  \label{fig:sdsssam}
\end{figure}
\subsection{Void galaxy properties}
Void galaxies do appear to have significantly different properties than average field galaxies. They appear to reside in a more 
youthful state of star formation and possess larger and less distorted supplies of gas. Analysis of void galaxies in the SDSS and 
2dFGRS indicate that void galaxies are bluer and have higher specific star formation rates than galaxies in denser 
environments \cite{groggell1999,groggell2000,rojas2004,rojas2005,patiri2006a,park2007,bendabeck2008,vogeley2009}. 

While these trends appear to be quite general, controversy persists in the literature as to whether or not galaxies in voids 
differ in their internal properties from similar objects in denser regions. Part of this relates to the selection of 
void galaxies. There are distinct differences between the majority of these galaxies, residing in the moderate 
underdense environment near the boundary, and those in the deepest realms, $\delta < -0.8$, where some 10\% of the 
void galaxies resides. The most recent results on the basis of SDSS DR6 \cite{vogeley2009} and 2dFGRS \cite{bendabeck2008} 
do indicate that these are intrinsic differences: at fixed morphology 
and luminosity the faintest void galaxies are intrinsically bluer and have higher star formation rates. It may certainly be true 
that, following the interpretation by ref.~\refcite{bendabeck2008}, because structure grows more slowly in the low-density 
environment of voids the galaxy population in voids is younger. An additional effect may be that star formation events 
may be triggered more easily by galaxy encounters in the low velocity dispersion environment of voids \cite{groggell1999}.

Interestingly, voids also contain nearly normal ellipticals \cite{park2007,croton2008}. It seems that the red sequence of ellipticals 
in the color-magnitude diagram from SDSS is almost independent of environment \cite{park2007}, contrasting sharply to the 
significant blueward shift of the blue sequence in voids \cite{blanton2005,park2007}. If major mergers produce ellipticals, 
it is not obvious how galaxies in the least dense regions, where interactions are rare, acquire these properties. Careful 
examination of ellipticals in voids may therefore reveal important clues to the balance of nature vs. nurture in galaxy formation. 

\begin{figure}[b]
  \centering
  \includegraphics[width=1.0\linewidth]{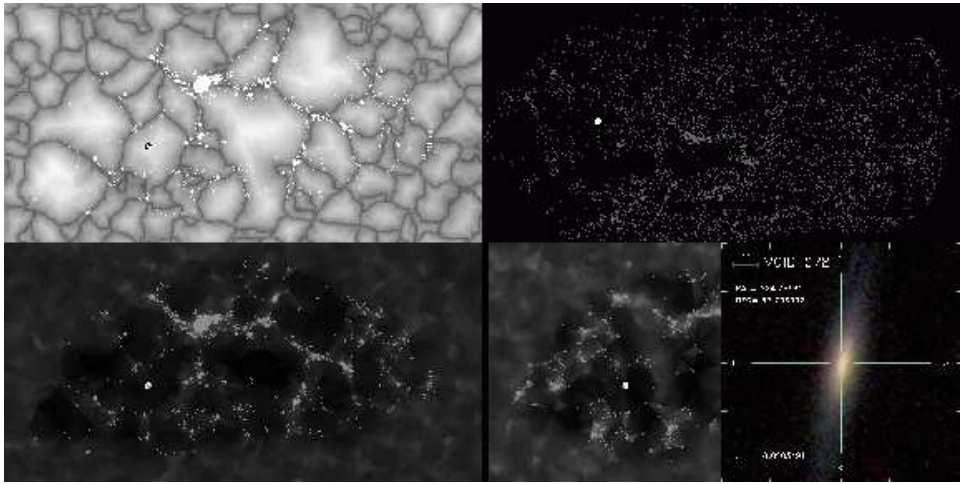}
  \caption{\small The SDSS void galaxy selection procedure. Shown is the location of galaxy void078 in 
our sample. Top righthand: the SDSS galaxy sky map (z $<$ 0.03), with sky position of the galaxy (white 
dot). The 3-D galaxy sample is translated into 
a spatial DTFE density map, (lower lefthand and centre), and a cosmic spine map (top lefthand). 
Lower lefthand and centre: spatial position of the void galaxy candidate (white dot) with respect to its 
embedding void, seen from two mutually perpendicular directions. The spatial information helps to trace 
the deepest void galaxies. Image courtesy Miguel Arag\'on-Calvo.}
  \label{fig:sdsssel}
\end{figure}
\subsection{Gas content of void galaxies: the WSRT HI void galaxy survey}
Even though a crucial factor for understanding their recent formation and star formation history, less is known about 
the gas content of void galaxies. Possible systematic trends in the ratio of hydrogen mass to stellar mass are crucial 
for understanding the influence of the void environment on the galaxies' star formation history. The morphology of 
the gas layer, in particular that of the outskirts, may reflect their isolated existence.  

The earliest systematic survey of HI in void galaxies is that of the galaxies in the Bo\"otes region \cite{szomoru1996}. It comprised 
pointed observations of 24 IRAS selected galaxies. Sixteen were detected with sufficient amounts of hydrogen, most of which were 
late-type, gas-rich and usually disturbed systems. Interesting was the finding of 18 new galaxy companions in the vicinity of the known 
galaxies. Nonetheless, most of the observed galaxies were residing in the outer realms of the void and with some reason might 
have been identified with the moderate density environment of walls. An important additional HI study of void galaxies involved the 
search for dwarf galaxies in the Local Void \cite{karachentsev2004}. The fact that they were only found at the edge of the void and 
avoid its interior represents a major challenge for our understanding of galaxy formation. Some well-known 
nearby luminous galaxies whose HI layers have been studied in detail, ie. NGC6946 and M101, are actually found within the 
Local Void. It is rather intriguing that e.g. NGC6946 is known to undergo a substantial level of gas inflow and has a large 
gas disk bearing the marks of numerous recent supernovae \cite{boomsmaphd2007}. Galaxies in and around voids are also searched 
for in the HIPASS \cite{kraankorteweg2007} and ALFALFA survey \cite{giov2008}. 

\begin{figure}[t]
  \centering
  \includegraphics[width=13.0cm]{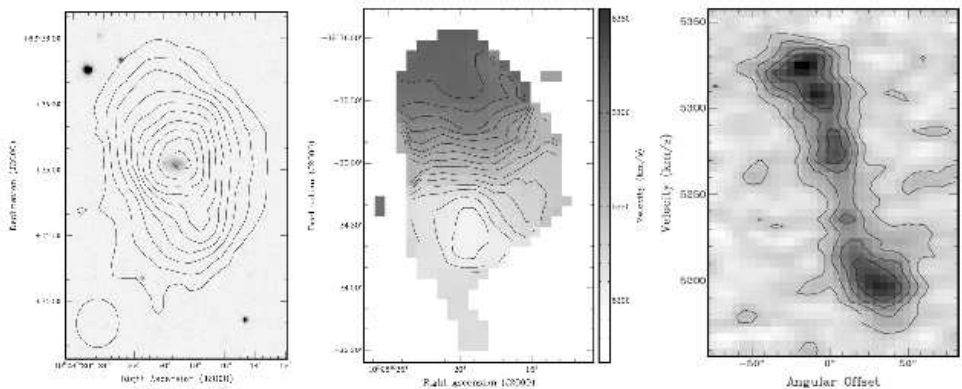}
  \caption{\small The polar ring galaxy J102819 23+623502.6 (void03). It is one of the most lonely 
galaxies in the Universe, with no NGC and UGC galaxy within 4 Mpc ! Shown are the 
HI intensity map superimposed on SDSS g band optical image (left), the velocity map 
(centre) and the position-velocity diagram along the major axis. The total HI mass 
in the polar ring is $2.0\times10^9 M_\odot$. See Stanonik et al. 2009a.}
  \label{fig:void03}
\end{figure}

The most sensitive and systematic study of gaseous void galaxies is our (ongoing) WSRT HI survey of SDSS galaxies. 
It consists of a sample of around 55 galaxies, sufficiently large to probe the morphological diversity of 
void galaxies and voids of different size, shape and underdensity.

It involves a unique selection procedure in that it is based upon finding void galaxies in the deepest underdensities 
of the DTFE reconstruction of the density field (DTFE: Delaunay Tessellation Field Estimator) in the (DR7) SDSS redshift survey 
\cite{schaapwey2000,schaapphd2007,weyschaap2009}. Fig.~\ref{fig:sdsssel} illustrates the void galaxy selection pipeline. 
Each sample galaxy is optically selected from SDSS/DR7 and has to be located in the central interior region 
of a nearby void identified from the SDSS survey. Al sample galaxies are at a considerably closer distance than the 
Bo\"otes void. Most galaxies in our sample are blue, a few red (see 10 specimen fig.~\ref{fig:sdsssam}). 

Results of the pilot study are reported in ref.~\refcite{stanonik2009a} and 
ref.~\refcite{stanonik2009b}. With 14 detections out of of 15 void galaxies, the detection rate of the 
pilot survey has been 93\% \cite{stanonik2009b}. The one without a detection is the reddest galaxy in the sample. 
The HI content of the 14 detected galaxies ranges from $\sim 6.5 \times 10^7$ M$_{\odot}$ to $\sim 4.8 \times 10^9$ M$_{\odot}$. Their 
optical radii are small, 5-10 kpc, while they have HI disks that extend far beyond the optical and appear to be in regular rotation. 
Nearly 2/3rd appear asymmetric or slightly warped at the faintest contour levels. Three galaxies turn out to be uniquely 
interesting systems. One is part of a system of three galaxies, all aligned in the plane of the sky, stretching out over 57 kpc, and 
connected by an HI bridge possibly related to a tenuous void filament. A second galaxy shows regular disk rotation in its 
HI distribution. However, optically it exhibits a clumpy, knotty morphology similar to that seen in chain-galaxies 
forming at high redshift. It also has a companion. 

The most outstanding specimen of the sample is a polar disk galaxy \cite{stanonik2009a} (see fig.~\ref{fig:void03}). 
Amongst the most lonely galaxies in the universe (4 Mpc up to the nearest UGC and NGC galaxy), it has a massive, 
star-poor HI disk that is perpendicular to the disk of the central void galaxy. No optical counterpart to the HI disk has 
been found, even though the central galaxy is actively forming stars. The ring reveals a regular rotation pattern. 

The galaxy is located within a what looks like a tenuous wall in between two large roundish voids. The central optical disk 
is strongly inclined with respect to the surrounding wall ($\sim 60^{\circ}-70^{\circ}$). The unique void environment forces 
a reappraisal of the traditionally assumed formation mechanisms for such polar ring galaxies, by a close interaction or 
merger. Its existence may point to a slow HI accretion dictated by the unique environment, possibly at the crossing point 
of the outflow from the two voids.\\

\subsection{Missing Void Dwarfs: $\Lambda$CDM challenged ?}
With respect to their spatial distribution, void galaxies are found to display a significant luminosity 
dependence. The consensus is that there is a good statistical agreement between theory and observation of 
the large voids defined by $L^*$ galaxies \cite{hoyvog2004,conroy2005,hoeft2006,tinker2009}. The observed 
density contrast of bright galaxies in large voids, around 1/10 of the cosmic mean \cite{hoyvog2002b,hoyvog2004}, 
is consistent with theoretical predictions. 

However, conflicts arise when we look at the distribution and properties of lower luminosity/mass objects. 
From optical and HI surveys, ref.~\refcite{karachentsev2004} found that the density of faint ($-18<M_B<-12$) galaxies 
is only $1/100$ that of the mean. This forms a marked contrast to the prediction by high-resolution 
$\Lambda$CDM simulations that voids have a density of low mass ($10^9 \hbox{\rm M}_{\odot}< \hbox{\rm M}<10^{11} 
\hbox{\rm M}_{\odot}$) halos that is 1/10 that of the cosmic mean \cite{warren2006,hoeft2006,tikhonov2009}. 
 
The dearth of dwarf and/or low surface brightness void galaxies represents a major challenge for the standard 
cosmological view of galaxy formation. The simplest models of galaxy formation \cite{little1994} predict 
that voids would be filled with galaxies of low luminosity, or galaxies of some other uncommon nature \cite{hoffman1992}. 
Ref.~\refcite{peebles2001} pointed out that the observed salient absence of dwarf galaxies in nearby voids could possibly 
involve strong ramifications for the viability of the $\Lambda$CDM cosmology on small scales. Branded with the name 
\textit{void-phenomenon}, the issue has been marked as a {\it ``a crisis for the CDM model''} since Peebles strongly 
emphasized the notion that nearby voids do {\it NOT} appear to contain the expected population of dwarf and/or low 
surface brightness galaxies. 

Various processes and influences have been forwarded as an explanation for the missing void dwarfs within the 
context of existing galaxy formation theories. These involve environmental properties of dark 
matter halos and astrophysical procesess, ranging from supplementary baryonic physical processes and 
radiation feedback processes. 

The nature of the environmental influence on the formation of halos has been addressed in a variety of studies. 
High-resolution simulations now clearly predict dependence of the shape, concentration, spin, and mean age of formation 
of halos on their environment \cite{sheth2002,gao2005,wechsler2006,harker2006,aragon2007,hahn2007}, particularly so for 
lower mass halos. Because the mass function of halos shifts to lower masses in voids these effects should be most noticeable 
among void galaxies \cite{gottloeb2003,goldvog2004}. Ref.~\refcite{furlpir2006} and ref.~\refcite{tinker2009} modelled 
the dependence of the properties of galaxies on the halos they inhibit to infer the properties of voids in the galaxy distribution. 
According 
to ref.~\refcite{tinker2009}, halo models would explain away the void phenomenon. However, the implied spatial segregation between 
galaxies of different luminosities in and near voids does not bear much resemblance to reality. The issue still seems far 
from solved and progress will depend largely on insights offered by new observational data. 

Amongst the most plausible astrophysical explanations for the absence of dwarfs in voids is the ionizing influence of 
the UV background in suppressing the star forming capacity of low mass halos \cite{benson2003}. However, 
ref.~\refcite{hoeft2006} showed that these effects would not explain the distribution of dwarf galaxies in voids. Also the suggestion 
that supernova feedback would be instrumental in evicting gas from the shallow potential wells of the dwarfs does not 
seem to offer an explanation. Simulations by ref.~\refcite{tassis2008} showed that most internal properties of dwarf galaxies can be 
reproduced without invoking supernova feedback.

\section*{Acknowledgments}
Our voyage along the empty realms of our universe has been made possible by the many inspiring 
and instructive discussions and keen insights of those who shared the experience along minor 
or major parts of the journey. In particular Miguel Arag\'on-Calvo, Joerg Colberg, Jacqueline van Gorkom, 
Thijs van der Hulst, Vincent Icke, Bernard Jones, Eelco van Kampen, Mark Neyrinck, Jim Peebles, George Rhee, 
Willem Schaap, Sergei Shandarin, Ravi Sheth, Kathryn Stanonik, Esra Ti{\u{g}}rak and Michael Vogeley, along with 
Jounghee Lee, have been instrumental in this quest for the {\it void}. RvdW is particularly grateful to Changbom Park 
for the invitation to this wonderful and interesting conference and for the great hospitality at KIAS where 
preparations for this paper commenced.

\end{document}